\documentclass[11pt]{article}
\newlength{\vshift}
\newlength{\hshift}
\usepackage{amssymb,amsopn}

\def\la{\lambda}

\def\de{\delta}

\def\al{\alpha}

\def\ds{\stackrel{\star}{,}}
\def\x{\hat x}

\def\p{\partial}

\def\hp{\hat{\partial}}
\def\h{\hat}
\def\lb{\lbrack}
\def\rb{\rbrack}
\def\pat{\partial}
\begin{document}
\begin{titlepage}
\rightline{LMU-TPW 2003-08}
\rightline{MPP-2003-29}

\vspace{4em}
\begin{center}

{\Large{\bf Deformed Field Theory on $\kappa$-spacetime}}

\vskip 3em

{{\bf Marija Dimitrijevi\' c${}^{1,3}$, Larisa Jonke${}^{1,4}$,
    Lutz M\"oller${}^{1,2}$,\\ Efrossini Tsouchnika${}^{1}$, Julius Wess${}^{1,2}$, Michael Wohlgenannt${}^{1}$  }}

\vskip 1em

${}^{1}$Universit\"at M\"unchen, Fakult\"at f\"ur Physik\\
        Theresienstr.\ 37, D-80333 M\"unchen\\[1em]

${}^{2}$Max-Planck-Institut f\"ur Physik\\
        F\"ohringer Ring 6, D-80805 M\"unchen\\[1em]

${}^{3}$University of Belgrade, Faculty of Physics\\
Studentski trg 12, SR-11000 Beograd\\[1em]

${}^{4}$Rudjer Boskovic Institute, Theoretical Physics Division\\
PO Box 180, HR-10002 Zagreb\\[1em]
 \end{center}

\vspace{2em}

\begin{abstract} 
A general formalism is developed that allows the construction of a
field theory on quantum spaces which are deformations of ordinary
spacetime. The symmetry group of spacetime
(Poincar\' e group) is replaced by a quantum group. This
formalism is demonstrated for the $\kappa$-deformed Poincar\'e algebra and
its quantum space. The algebraic setting is mapped to the algebra of
functions of commuting variables with a suitable
$\star$-product. Fields are elements of this function algebra. The
Dirac and Klein-Gordon equation are defined and an action is found
from which they can be derived.
\end{abstract}
\vskip 1.5cm
\qquad\hspace{2mm}\scriptsize{eMail: dmarija; larisa; lmoeller;
frosso; wess; miw@theorie.physik.uni-muenchen.de}
\vfill

\end{titlepage}\vskip.2cm

\newpage
\setcounter{page}{1}
\section{Introduction}

All experimental evidence supports the assumption that space
time forms a differential manifold. \\
All successful fundamental
theories are formulated as  field theories 
on such manifolds. 

Nevertheless, in quantum field theories (QFT) we meet some intrinsic
difficulties at very high energies or very short distances that do
not seem to resolve in the framework of QFT. It seems that 
the structure of QFT has to be modified somewhere. We have
no hints from experiments where and how this should be done.

In a very early attempt - almost at the beginning of QFT - it was
suggested by W.~Heisenberg \cite{heisenberg} that spacetime might be modified
at very short distances by algebraic properties that could lead
to uncertainty relations for the space coordinates.

This idea was worked out by H. S. Snyder \cite{snyder} in a specific
model. He gave a very systematic analysis and physical
interpretation of such a structure. W.~Pauli, in a letter to
N.~Bohr \cite{pauli} called it ``a mathematically ingenious proposal,
which, however, seems to be a failure for reasons of physics''. 

In the meantime experimental data for physics at much
shorter distances are available. At the same time 
mathematical methods have improved enormously and it
seems to be time to exploit the idea again.

In mathematics the concept of deformation has shown to be 
extremely fruitful. Especially the deformation of groups
to Hopf algebras \cite{jimbo}, \cite{drinfeld}, \cite{woronowicz} and \cite{faddeev}, the so called quantum groups, has
opened a new field in mathematics. At the same time
the deformation of quantum mechanics \cite{flato} has seen
a very exciting development as well. 

In this paper we try to bring these two 
concepts together aiming at a 
Deformed Field Theory (DFT). It is not a differential
manifold on which we formulate such a 
theory, it is rather formulated on a
quantum space. 

An example is the canonical quantum space where the
coordinates $\hat{x}^\mu$ are subject to the relations
\begin{displaymath}
[\hat{x}^\mu,\hat{x}^\nu]=\theta^{\mu\nu},
\end{displaymath}
with constant $\theta^{\mu\nu}$. This structure has
been investigated in many papers (e.g. \cite{madore} and the references in
\cite{szabo} and \cite{nekrasov}). There is, however,
no quantum group associated with this quantum space. 

We expect additional features of a field theory from a quantum group 
that can be interpreted as a deformation of the 
Poincar\' e group. The simplest example is the $\kappa$-deformed
Poincar\'e  algebra and its associated quantum space. In this paper we treat the
Euclidean version:
\begin{displaymath}
[\hat{x}^\mu,\hat{x}^\nu]=i(a^\mu \h{x}^\nu-a^\nu\h{x}^\mu),\qquad
\mu=1,\dots n.
\end{displaymath}

The algebraic 
structure of the $\kappa$-deformed Poincar\' e algebra has been
investigated intensively, e.g. \cite{lukrue1}, \cite{lukrue2}, \cite{majrue} and \cite{kosinski}. 

In this paper we have systematically developed the
approach starting from the coordinate algebra in the spirit of
Y.~Manin's discussion of $SU_q(2)$ \cite{Manin}. In this approach, the
coordinate algebra becomes a factor space and all maps on this space
have to respect the factorization property. We then use the 
isomorphism of the 
abstract algebra with the algebra of functions of
commuting variables equipped with a $\star$-product. In series of
papers by J.~Lukierski et al. \cite{local}, \cite{wigner} and
\cite{klm0307} (see also \cite{gac1}),
the model has been treated with the methods of deformation quantization
as well. Their work is very similar to our approach. 

In the second chapter we concentrate on the $\kappa$-deformed
quantum space in the algebraic setting. 
We define derivatives, generators of the deformed symmetry  algebra,
as well as Dirac and Laplace operators, constructing the model
systematically on the basis of the quantum space. All formulae
are worked out in full detail.

In the third chapter we map the algebraic setting into
the framework of deformation quantization, introducing a suitable $\star$-product. 

In the fourth chapter we introduce fields that are going to be the
objects in a DFT. The Klein-Gordon equation and
the Dirac equation are formulated.

In the last chapter we introduce an integral for an action. Field
equations can then be derived by means of a variational
principle.

\section{The Algebra}

\vspace*{0.3cm}
The symmetry structure of  $\kappa$-Minkowski spacetime is an example of a quantum
group (Hopf algebra) that acts on a quantum space (module). Our aim
is to construct quantum field theories with
methods of deformation quantization on such a quantum
space and to study the implications of a quantum group
symmetry on  these field theories.

For this purpose we start from the quantum
space and the relations that define it.

\vspace*{0.5cm}
\noindent
{\bf Coordinate Space}

\vspace*{0.3cm}

The coordinate space will be the factor space
of the algebra freely generated by the 
coordinates $\hat{x}^1\dots \hat{x}^n$,
divided by the ideal generated by commutation relations (cp.
\cite{calculus}, \cite{Woronowicz2} and \cite{Manin}). For the $\kappa$-Minkowski space the relations
are of the Lie algebra-type\footnote{The deformation parameter $a^\mu$ is related to the more
  common $\kappa$ through $\sqrt{a^2}=\kappa^{-1}$.}
\begin{equation}
\lbrack \h{x}^\mu ,\h{x}^\nu\rbrack = i(a^\mu \h{x}^\nu-a^\nu\h{x}^\mu),\quad \mu=1,\dots n. \label{1.1}
\end{equation}
The real parameters $a^\mu$ play the role of structure
constants for the Lie algebra:
\begin{equation}
\lbrack \h{x}^\mu ,\h{x}^\nu\rbrack = iC^{\mu\nu}_\rho \h{x}^\rho, \qquad C^{\mu\nu}_\rho = a^\mu \de^\nu_\rho -a^\nu \de^\mu_\rho.
\label{1.1a}
\end{equation}

We here study the Euclidean version. The generalization to a
Poincar\' e version is straightforward, with the direction $n$ (see
(\ref{1.2})) either space-, time- or light-like\footnote{Compare
\cite{lukruegg}, \cite{ball1} and \cite{klm0307}.}. In the Euclidean case $a^\mu$ can be
transformed by a linear transformation of the coordinates
(rotation) to the form:
\begin{equation}
\label{1.2}
a^\mu=\delta ^{\mu n}a. 
\end{equation}
The vector $a^\mu$ points into the $n$ direction.
In this form it is easier to analyse the relations (\ref{1.1}), they
are
\begin{eqnarray}
\label{1.3}
\lbrack \h{x}^i,\h{x}^j\rbrack &=& 0,  \nonumber \\
\lbrack \h{x}^n,\h{x}^i\rbrack &=& ia\h{x}^i,\quad  i,j=1,\dots n-1. 
\end{eqnarray}

\vspace*{0.5cm}
\noindent
{\bf {\bf $SO_a(n)$} Rotations}

\vspace*{0.3cm}

A map of the coordinate space has to
respect the factor space structure, or as we say it has to be
consistent with the relations (cp. \cite{Manin} and \cite{calculus}). Generators of
such maps are
\begin{eqnarray}
\label{1.4}
\lb M^{rs},\hat x^i\rb &=& \delta^{ri} \hat x^s - \delta^{si} \hat x^r ,\nonumber\\
\lb M^{rs}, \hat x^n\rb &=& 0, \nonumber\\
\lb N^l, \hat x^i\rb &=& -\delta^{li} \hat x^n -ia M^{li}, \\
\lb N^l, \hat x^n\rb &=& \hat x^l + iaN^l. \nonumber
\end{eqnarray}
We shall call $M^{rs}$  and $N^l=M^{nl}$ generators of
$SO_a(n)$, because for $a=0$ we find the generators of the rotation group
$SO(n)$. For $a\neq 0$ we have to check  the consistency
of (\ref{1.3}) and  (\ref{1.4}). Since this
type of calculations will appear again and again, we
exhibit an example. We calculate
\begin{equation}
N^l\Big( \lb \h{x}^n,\h{x}^i\rb-ia\h{x}^i\Big) \label{1.5} 
\end{equation}
term by term using (\ref{1.4}):
\begin{eqnarray}
N^l\x ^n\x ^i &=& \x ^l\x ^i-ia(\delta^{il}\x ^n+iaM^{li})+ia\x^i N^l \nonumber \\
&&-\delta^{il}\x ^n\x ^n -ia\x ^nM^{li}+\x ^n\x ^iN^l, \nonumber \\
N^l\x ^i\x ^n &=& -\delta^{il}\x ^n\x ^n -ia\x ^nM^{li}+\x ^i\x ^l+ia\x ^iN^l+\x ^i\x ^nN^l, \nonumber \\
N^l(-ia\x ^i) &=& ia(\delta^{il}\x ^n+iaM^{li}-\x ^iN^l) .\nonumber
\end{eqnarray}
Adding all up we find
\begin{equation}
N^l\Big( \lb \h{x}^n,\h{x}^i\rb-ia\h{x}^i\Big)=\Big( \lb \h{x}^n,\h{x}^i\rb-ia\h{x}^i\Big)N^l. \label{1.6}
\end{equation}
The consistency of the $N^l$ operations with (one of) the relations (\ref{1.3})
is verified. 

If we now define the map
\begin{equation}
\h{x}'^\mu = \h{x}^\mu+ \epsilon_l \left( N^l \h{x}^\mu\right),
\label{1.6a}
\end{equation}
we find to first order in $\epsilon$:
\begin{equation}
[\h{x}'^n,\h{x}'^i] = ia\h{x}'^i,\qquad [\h{x}'^i,\h{x}'^j]=0 .
\label{1.6b}
\end{equation}

The algebra
generated by the
rotations  is a deformation of the Lie algebra $SO(n)$,
we shall call it $SO_a(n)$.

From (\ref{1.4}) it is possible to compute the commutators
of the
generators. As a possible solution (this was considered specifically
in \cite{majrue}) we find the undeformed $SO(n)$
algebra:
\begin{eqnarray}
\label{1.7}
\lb N^l,N^k\rb &=& M^{lk}, \nonumber \\
\lb M^{rs},N^l\rb &=& \delta^{rl}N^s - \delta^{sl}N^r, \\
\lb M^{rs},M^{kl}\rb &=& \delta^{sl} M^{rk} + \delta^{rk}M^{sl}-\delta^{rl}M^{sk}  
- \delta^{sk} M^{rl}. \nonumber
\end{eqnarray}
But the comultiplication
will turn out to
be quite different when the generators act on functions of
$\hat{x}$. This is already apparent from (\ref{1.4}).
An explicit expression for the comultiplication will contain
derivatives as well. Thus, we define derivatives next.

%
%
%
%
%
%

\vspace*{0.5cm}
\noindent
{\bf Derivatives}

\vspace*{0.3cm}

Derivatives on an algebra have been introduced in \cite{calculus}. They
generate a map in the coordinate space, 
elements of the coordinate space are mapped to other 
elements of the coordinate space. Thus, they have to be consistent
with the algebra relations. For $a =0$ they should behave
like ordinary derivatives, for $a \neq 0$
the Leibniz rule has  to be generalized to achieve consistency  \cite{calculus}.

We also demand that the derivatives form a
module for $SO_a(n)$. In addition they should act 
at most linearly in the
coordinates and  the derivatives. These requirements are
satisfied by the following rules for differentiation:
\begin{eqnarray}
\label{1.8}
\lb\hp _n, \hat x^i\rb &=& 0,\nonumber\\
\lb\hp _n, \hat x^n\rb &=& 1,\nonumber\\
\lb\hp _i, \hat x^j\rb &=& \delta^j_i,\\
\lb\hp _i, \hat x^n\rb &=& ia \hp _i \nonumber
\end{eqnarray}
and
\begin{equation}
\lb \hp _\mu,\hp _\nu \rb =0. \label{1.9}
\end{equation}
The requirement of linearity has been added in order to
get an (almost) unique solution\footnote{see \cite{sit}, \cite{lukzak} and \cite{forthcoming}.}. It is not essential for the
definition of derivatives. 

We can apply derivatives to a function of $\hat{x}^\mu$
and take the derivatives to the right hand side of this function 
using (\ref{1.8}). For the $\hat{\partial}_n$
this yields the usual Leibniz rule, for the $\hat{\partial}_i$
we find that $\hat{x}^n$ is shifted by $ia$.
This can be expressed by the shift operator $e^{ia\hat{\partial}_n}$.
Note that $\hat{\partial}_n$ commutes with $\hat{x}^i$.

We obtain the Leibniz rule:
\begin{eqnarray}
\h{\partial}_n(\hat{f}\cdot \hat{g})&=& (\h{\partial}_n\hat{f})\cdot \hat{g}+
\hat{f}\cdot \h{\partial}_n\hat{g},\nonumber \\
\h{\partial}_i(\hat{f}\cdot \hat{g})&=& (\h{\partial}_i\hat{f})\cdot \hat{g}
+(e^{ia\hp_n}\h{f})\cdot \h{\partial}_i\hat{g} .\label{1.8a}
\end{eqnarray}

Next we construct commutators of the generators of $SO_a(n)$ with the
derivatives such that these form a module. For
this purpose we perform a power series expansion in $a$,
at lowest order we start from a vector-like behavior
of $\hat{\partial}_\mu$. In first order in $a$ we have to 
modify the commutator to be consistent with
(\ref{1.7}) and (\ref{1.8}). This procedure has to be repeated, finally
we find
\begin{eqnarray}
\label{1.10}
\lb M^{rs}, \hp _i\rb &=& \delta ^r_i \hp_s -  \delta ^s_i \hp_r, \nonumber\\
\lb M^{rs}, \hp _n\rb &=& 0, \nonumber\\
\lb N^l, \hp _i\rb &=& \delta ^l_i\frac{1-e^{2ia\hp _n}}{2ia}  -
\frac{ia}{2}\delta^ l_i \hat{\Delta}+ ia\hp _l\hp _i, \\
\lb N^l, \hp _n\rb &=& \hp_l, \nonumber
\end{eqnarray}
where $\hat{\Delta}=\sum_{i=1}^{n-1} \hp _i\hp _i$. Equations (\ref{1.10})
are valid to all orders in $a$. By a direct construction we
have shown that the derivatives form a module of $SO_a(n)$.

We can apply $M^{rs}$ and $N^l$ to a function of $\hat{x}^\mu$
and take the generators to the right hand side. From the result of 
this calculation we can abstract the comultiplication rule: 
\begin{eqnarray}
\label{1.11}
\Delta N^l &=& N^l\otimes{\mathbf 1} + e^{ia\hp _n}\otimes N^l -
ia\hp _b\otimes M^{lb}, \nonumber\\
\Delta M^{rs} &=& M^{rs}\otimes{\mathbf 1} +  {\mathbf 1}\otimes M^{rs}, \nonumber \\
\Delta\hp _n &=& \hp _n \otimes {\mathbf 1} + {\mathbf 1}\otimes\hp _n, \\
\Delta\hp _i &=& \hp _i \otimes {\mathbf 1} + e^{ia \hp _n}\otimes \hp _i. \nonumber
\end{eqnarray}
This is the comultiplication consistent with the algebra
(\ref{1.7}). It  is certainly different from the comultiplication rule
for $SO(n)$. 

The comultiplication involves the derivatives.
For representations of the
algebra (\ref{1.7}), where $\hat{\partial}_\mu$ acting
on the representation gives zero, 
the standard comultiplication
rule for $SO(n)$ emerges.

As far as the commutators of $M^{rs}$ and $N^l$ with the
coordinates and the derivatives are concerned, we can
express $M^{rs}$ and $N^l$ by the coordinates and the derivatives,
as it is usually done for angular momentum:
\begin{eqnarray}
\label{1.11a}
\hat{M}^{rs} &=&\hat{x}^s\hat{\partial}_r -\hat{x}^r\hat{\partial}_s,\nonumber\\
\hat{N}^{l} &=&\hat{x}^l\frac{e^{2ia\hat{\partial}_n}-1}{2ia}-\hat{x}^n\hat{\partial}_l+\frac{ia}{2}\hat{x}^l\hat{\Delta}.
\end{eqnarray}

According to (\ref{1.11}), it
is natural to consider the generators
$M^{rs}$, $N^l$ and $\hat{\partial}_\mu$ as generators of
the $a$-Euclidean Hopf algebra. It should be noted that the deformed generators $M^{rs}$, $N^l$ do not
form a Hopf algebra by themselves. In the coproduct the
derivatives, or equivalently the translations in the
$a$-Euclidean Hopf algebra, appear as well.

%
%
%
%
%
%

\vspace*{0.5cm}
\noindent
{\bf Laplace and Dirac operators}

\vspace*{0.3cm}

A deformed Laplace operator (see \cite{lukrue1}, \cite{lukrue2}) and a deformed
Dirac operator (see \cite{Dirac1}, \cite{Dirac2}) can be defined. For
the Laplace operator $\hat \Box$  we demand that it commutes
with the generators of the $a$-Euclidean Hopf algebra
\begin{equation}
\lb M^{rs},\hat{\Box}\rb =0, \quad\quad \lb N^{l},\hat{\Box}\rb =0,\label{1.12a}
\end{equation}
and that it is a deformation of the usual Laplace
operator. By iteration in $a$ we find\footnote{In this form the
Laplace operator has 
been given in \cite{kosinski}.}:
\begin{equation}
\hat \Box = e^{-ia\hp _n} \hat{\Delta} +{2 \over a^2}
\Big( 1 - \cos ( a\hp _n ) \Big). \label{1.12}
\end{equation}

Since the $\gamma$-matrices are $\hat{x}$-independent and
transform as usual, the covariance of the full Dirac operator
$\gamma^\mu \hat{D}_\mu$ implies
that the transformation law of its components is vector-like:
\begin{eqnarray}
\lb M^{rs},\hat{D}_n\rb &=& 0, \nonumber \\
\lb M^{rs},\hat{D}_i\rb &=& \delta _i^r\hat{D}_s-\delta _i^s\hat{D}_r, \nonumber \\
\lb N^l,\hat{D}_n\rb &=& \hat{D}_l, \label{1.13}\\
\lb N^l,\hat{D}_i\rb &=& -\delta _i^l\hat{D}_n. \nonumber
\end{eqnarray}
These relations are obviously consistent with the algebra
(\ref{1.7}). 
A differential operator that satisfies
(\ref{1.13}) and that has the correct limit for $a \rightarrow 0$ is:
\begin{eqnarray}
\hat{D}_n &=& {1\over a} \sin (a\hp _n)+
{ia\over 2}\hat{\Delta}\hspace{1mm} e^{-ia\hp _n}, \nonumber \\
\hat{D}_i &=&\hp _i e^{-ia\hp _n}, \label{1.14}
\end{eqnarray}
where the derivatives $\hat{\partial}_\mu$ transform according
to (\ref{1.10}). 

The square of the Dirac operator turns out to be (compare \cite{Dirac1}):
\begin{equation}
\gamma^\mu\hat{D}_\mu\gamma^\nu\hat{D}_\nu=\sum _{\mu=1}^n \hat{D}_\mu\hat{D}_\mu = \hat \Box \Big( 1-{a^2\over 4}\hat\Box
\Big) .\label{1.15}
\end{equation}
Using this we can express the Laplace operator as
a function of the Dirac operator:
\begin{equation}
\hat{\Box}={2\over a^2}\Big( 1-\sqrt{1-a^2{\hat D}_\mu{\hat D}_\mu}\Big) .\label{1.16}
\end{equation}
The sign of the square root is determined by the limit
$a\rightarrow 0$. We have dropped the summation symbol.

\vspace*{0.5cm}
\noindent
{\bf Dirac operator as a derivative}

\vspace*{0.3cm}

The Dirac operator $\hat{D}_\mu$ can be seen as a derivative
operator as well,
having very simple transformation properties under
$SO_a(n)$, but as we shall see with a highly non-linear Leibniz rule.

We invert (\ref{1.14}) in order to express
the derivative operator $\hat{\partial}_\mu$ in terms
of the Dirac operator and   proceed as follows:
\begin{eqnarray}
\hp _i &=& \hat{D}_ie^{ia\hp _n}, \nonumber \\
\hp _i\hp _i &=& \hat{\Delta}=\hat{D}_i\hat{D}_ie^{2ia\hp _n},\nonumber\\
\hat{D}_n &=& {1\over 2ia}\Big( e^{ia\hp _n}-e^{-ia\hp _n}\Big) +{ia\over 2}
\hat{D}_i\hat{D}_ie^{ia\hp _n}. \label{1.17}
\end{eqnarray}
Multiplying equation (\ref{1.17}) by $e^{-ia\hp _n}$ leads to a quadratic
equation for $e^{-ia\hp _n}$:
$$
e^{-2ia\hp _n}+2ia\hat{D}_ne^{-ia\hp _n}+a^2\hat{D}_i\hat{D}_i-1=0 .
$$
Solving this quadratic equation we find (compare \cite{DSR}):
\begin{equation}
e^{-ia\hp _n}=-ia{\hat D}_n+\sqrt{1-a^2{\hat D}_\mu{\hat D}_\mu} .\label{1.18}
\end{equation}
The sign of the square root is again determined by the limit
$a \rightarrow 0$.

With (\ref{1.16}) we find a  form
that is
easier to handle
\begin{equation}
e^{-ia\hp _n}=1-ia{\hat D}_n-{a^2\over 2}\hat{\Box} .\label{1.19}
\end{equation}

Multiplying equation (\ref{1.17}) by $e^{ia\hp _n}$ we
find a quadratic equation for $e^{ia\hp _n}$ with the
solution:
\begin{eqnarray}
e^{ia\hp _n} &=& {1\over 1-a^2\hat{D}_k\hat{D}_k}\left( ia\h{D}_n+\sqrt{1
-a^2{\hat D}_\mu{\hat D}_\mu}\right) \nonumber \\
&=& {1\over 1-a^2\hat{D}_k\hat{D}_k}\left( 1
+ia{\hat D}_n-{a^2\over 2}\hat{\Box} \right) . \label{1.20}
\end{eqnarray}
It is easy to verify that (\ref{1.20}) is the inverse
of (\ref{1.18}).

Now we can invert (\ref{1.14}):
\begin{eqnarray}
\hp _i &=& {\h{D}_i\over 1-a^2\h{D}_k\h{D}_k}\left( ia\h{D}_n +\sqrt{1-a^2{\hat D}_\mu{\hat D}_\mu}\right)
\nonumber \\
&=& {\h{D}_i\over 1-a^2\h{D}_k\h{D}_k}\left( 1
+ia{\h D}_n-{a^2\over 2}\h{\Box} \right) , \label{1.21}\\
\hp _n &=& -{1\over ia}\ln \left( -ia\h{D}_n +\sqrt{1-a^2{\hat D}_\mu{\hat D}_\mu}\right)=
-{1\over ia}\ln \left(1-ia{\h D}_n-{a^2\over 2}\h{\Box}\right) .\nonumber
\end{eqnarray}

To compute the commutator of $\hat{D}_\mu$ and $\hat{x}^\nu$,
we  use the representation (\ref{1.14}), apply (\ref{1.8}) and  finally
express $\hat{\partial}_\mu$ again by $\hat{D}_\mu$
using (\ref{1.21}). The result is
\begin{eqnarray}
\lb \h{D}_n, \hat{x}^j\rb  &=& ia\h{D}_j , \nonumber \\
\lb \h{D}_n, \hat{x}^n\rb &=& \sqrt{1-a^2{\hat D}_\mu{\hat D}_\mu} ,\nonumber \\
\lb \h{D}_i, \hat{x}^j\rb &=& \delta _i^j\left(
-ia{\hat D}_n+\sqrt{1-a^2{\hat D}_\mu{\hat D}_\mu}\right) , \label{1.22}\\
\lb \h{D}_i, \hat{x}^n\rb &=& 0.\nonumber
\end{eqnarray}

For two functions $\hat{f}(\hat{x})$ and $\hat{g}(\hat{x})$ the
Leibniz rule  can
be computed from (\ref{1.22})
\begin{eqnarray}
\h{D}_n(\hat{f}\cdot \hat{g}) &=& (\h{D}_n\hat{f})\cdot (e^{-ia\hp _n}\hat{g})+
(e^{ia\hp _n}\h{f})\cdot (\h{D}_n\hat{g})+ia(\h{D}_je^{ia\hp _n}\hat{f})\cdot (\h{D}_j\hat{g}) ,\nonumber \\
\h{D}_i(\hat{f}\cdot \hat{g}) &=& (\h{D}_i\hat{f})\cdot (e^{-ia\hp _n}\hat{g})
+\h{f}\cdot (\h{D}_i\hat{g}) .\label{1.23}
\end{eqnarray}
For $e^{\pm ia\hp _n}$ the expressions (\ref{1.18}) and (\ref{1.20}) have to be
inserted.

Equation (\ref{1.14}) tells us that the Dirac operator $\hat{D}_\mu$
is in the enveloping algebra of $\hp _\mu$ and (\ref{1.21})
that the derivative operator $\hp _\mu$ is in the enveloping
algebra of $\hat{D}_\mu$. Equations (\ref{1.14}) and (\ref{1.21}) can
be interpreted as a change of the basis in the derivative
algebra (compare \cite{DSR}).

One basis $\{\hat{D}_\mu\}$ has simple transformation properties,
the other basis $\{\hp _\mu\}$ has a simple Leibniz rule.

The $a$-Euclidean Hopf algebra might also be generated by $M^{rs}$, $N^l$ and 
$\hat{D}_\mu$. This will be of advantage if we focus on the
$SO_a(n)$ behavior:
\begin{eqnarray}
\lb M^{rs},M^{tu} \rb &=& \delta^{rt}M^{su}+\delta^{su}M^{rt}-\delta^{st}M^{ru}-\delta^{ru}M^{st}, \nonumber \\
\lb M^{rs},N^l \rb &=& \delta^{rl}N^s- \delta^{sl}N^r,\nonumber \\
\lb N^k,N^l \rb &=& M^{kl},\\
\lb M^{rs},\h{D}_n \rb &=&0,\nonumber  \\
\lb M^{rs},\h{D}_i \rb &=&\delta^{ri}\h{D}_s-\delta^{si}\h{D}_r, \nonumber \\
\lb N^l,\h{D}_n \rb &=&\h{D}_l, \nonumber \\
 \lb N^l,\h{D}_i \rb &=&-\delta^{li}\h{D}_n, \nonumber \\
\lb \h{D}_\mu, \h{D}_\nu \rb&=&0.\nonumber
\label{1.23aa}
\end{eqnarray}
This again is the undeformed  algebra, $a$ does not appear.
Of course, the comultiplication in this basis depends on
$a$:
\begin{eqnarray}
\Delta M^{rs}&=& M^{rs} \otimes {\mathbf 1}  +{\mathbf 1} \otimes M^{rs},\nonumber \\
\Delta N^l&=& N^l \otimes {\mathbf 1}  +\frac{ ia\h{D}_n+\sqrt{1
-a^2{\hat D}_\mu{\hat D}_\mu}}{1-a^2\hat{D}_j\hat{D}_j}\otimes N^l\nonumber\\
&&-\frac{ia\hat{D}_k}{1-a^2\hat{D}_j\hat{D}_j}\left(ia\h{D}_n+\sqrt{1
-a^2{\hat D}_\mu{\hat D}_\mu}\right)\otimes M^{lk},\nonumber \\
\Delta \h{D}_n &=&  \h{D}_n \otimes\left( -ia\h{D}_n+\sqrt{1
-a^2{\hat D}_\mu{\hat D}_\mu}\right)+\frac{ia\h{D}_n+\sqrt{1
-a^2{\hat D}_\mu{\hat D}_\mu}}{1-a^2\hat{D}_j\hat{D}_j}\otimes \h{D}_n\nonumber\\
&& +ia {\hat{D}_k\over 1-a^2\hat{D}_j\hat{D}_j}\left( ia\h{D}_n+\sqrt{1
-a^2{\hat D}_\mu{\hat D}_\mu}\right)\otimes \hat{D}_k, \\
\Delta \h{D}_i &=& \h{D}_i \otimes\left( -ia\h{D}_n+\sqrt{1
-a^2{\hat D}_\mu{\hat D}_\mu}\right)+{\mathbf 1} \otimes \h{D}_i.\nonumber
\label{1.23a}
\end{eqnarray}
To define a Hopf algebra, multiplication and comultiplication are
essential\footnote{The additional ingredients for a Hopf algebra,
  counit and antipode, have been calculated, e.g. \cite{kosmas}.}.

%
%
%
%
%
%

\vspace*{0.5cm}
\noindent
{\bf Conjugation}

\vspace*{0.3cm}
All the relations that we have considered do not change under the
formal involution that we shall call conjugation:
\begin{eqnarray}
(\hat{x}^\mu)^+=\hat{x}^\mu, &\quad& \quad (\hp _\mu)^+=-\hp _\mu ,\nonumber\\
(M^{rs})^+=-M^{rs}, &\quad& \quad (N^l)^+=-N^l. \label{1.28}
\end{eqnarray}
The order of algebraic elements in the product has to be inverted
under conjugation.

It is easy to show that $\hat{N}^{l}$ and $\hat{M}^{rs}$ 
in formula (\ref{1.11a}) have the desired conjugation property
by conjugating $\hat{x}^\mu$ and $\hat{\partial}_\mu$.

%
%
%
%
%
%

\section{The $\star$-product}

Our aim is to formulate a field theory on the algebra
discussed in the previous section with the methods of
deformation quantization. The algebraic formalism is
connected with deformation quantization via the $\star$-
product. The idea in short is as follows:
We consider polynomials of fixed degree in the algebra
- homogeneous polynomials. They form a finite-dimensional
vector space. If the algebra has the Poincar\' e-Birkhoff-Witt
property, and all Lie algebras have this property, then
the dimension of the vector space of homogeneous polynomials
in the algebra is the same as for polynomials of
commuting variables. Thus, there is an isomorphism
between the two finite-dimensional  vector spaces. This
vector space isomorphism can be extended to an algebra
isomorphism by defining the product of polynomials of 
commuting variables by first mapping these polynomials back
to the algebra, multiplying them there and mapping the product 
to the space of polynomials of ordinary variables. The product we obtain 
that way is called $\star$-product. It is noncommutative and
contains the information about the product in the algebra.
The objects that we will identify with physical fields are
 functions. This is possible because the $\star$-product
of polynomials can be extended to the $\star$-product of
functions.

\vspace*{0.5cm}
\noindent
{\bf The $\star$-product for Lie algebras}

\vspace*{0.3cm}

There is a standard $\star$-product for Lie algebras 
\cite{Kathotia}. If $\hat{x}^\mu$ are the generators of
a Lie algebra such that:
\begin{equation}
\label{2.1}
\lb \x^\mu , \x^\nu \rb = i C^{\mu\nu}_\rho \x^\rho,
\end{equation}
then a $\star$-product can be computed with the help of the
Baker-Campbell-Hausdorff formula
\begin{equation}
\label{2.2}
e^{i\x^\nu p_\nu} e^{i\x^\nu q_\nu} = e^{i\x^\nu \{ p_\nu + q_\nu + {1\over 2} g_\nu(p,q) \}}.
\end{equation}
The exponential, when expanded, is always fully symmetric in the
algebraic elements $\hat{x}^\nu$. Therefore we call this
$\star$-product the symmetric $\star$-product. In the following we shall use the symmetric $\star$-product. It is
\begin{equation}
\label{2.3}
f\star g(z) = \lim_{x\to z \atop y \to z} \exp \left( {i\over 2} z^\nu g_\nu(i\partial_x,i\partial_y)
\right) f(x)g(y). 
\end{equation}

It can be applied to any Lie algebra, but in general there is
no closed form for $g_\nu(i\partial_x,i\partial_y)$.
It can, however, be computed in a power series expansion 
in the structure constants $C^{\mu \nu}_\rho$. We obtain 
\begin{equation}
\label{2.33a}
[x^\mu \ds x^\nu] = x^\mu \star x^\nu - 
x^\nu \star x^\mu = iC^{\mu \nu}_\rho x^\rho.
\end{equation}
All the consequences of the algebraic relation (\ref{2.1}) can
be derived from the $\star$-product.

If the algebra allows a conjugation, then the symmetric
$\star$-product has the
property:
\begin{equation}
\label{2.5}
\overline{f\star g} = \bar g \star \bar f.
\end{equation}
The bar denotes complex conjugation.

We have found a closed form for the symmetric $\star$-product
for the algebra (\ref{1.3})
\cite{forthcoming} (compare \cite{wigner}).
Using the abbreviations
\begin{equation}
\label{2.4b}
\pat_{x^n}={\pat\over \pat x^n},\quad\pat_{y^n}={\pat\over \pat y^n},\quad \pat_n={\pat\over \pat x^n}+{\pat\over \pat y^n }
\end{equation}
the $\star$-product takes the form
\begin{eqnarray}
\label{2.4a}
 f \star  g(z) &=& \lim_{x\to z \atop y \to z}\exp \Bigg( z^j \pat_{x^j}
   \left( {\pat_n \over \pat_{x^n}} e^{-ia\pat_{y^n}} {1-e^{-ia\pat_{x^n}}
   \over 1-e^{-ia\pat_n}}- 1 \right)\nonumber\\
  && \qquad\qquad + z^j \pat_{y^j}
   \left( {\pat_n \over \pat_{y^n}} {1-e^{-ia\pat_{y^n}}
   \over 1-e^{-ia\pat_n}}- 1 \right)
   \Bigg) f(x)g(y).
\end{eqnarray}
To second order in $a$ we obtain:
\begin{eqnarray}
\label{2.4}
f \star g\, (x) & = & f(x)g(x) + {ia\over 2} x^j \Big( \pat_n f(x) \pat_j g(x) 
        -  \pat_j f(x) \pat_n g(x) \Big) \nonumber\\
& & -{a^2\over 12}x^j \Big( \pat_n^2 f(x) \pat_j g(x)-\pat_j\pat_n f(x) \pat_n g(x)\nonumber\\
&  &\qquad\qquad -\pat_n f(x) \pat_j \pat_n g(x)+ \pat_j f(x) \pat_n^2 g(x)\Big)  \nonumber\\
&  & -{a^2\over 8} x^j  x^k\Big(\pat_n^2 f(x) \pat_j\pat_k g(x) - 2
        \pat_j \pat_n f(x) \pat_n\pat_k g(x)\\
&  &\qquad\qquad
        +\pat_j \pat_kf(x) \pat_n^2 g(x) 
        \Big) + \mathcal{O}(a^3). \nonumber
\end{eqnarray}

%
%
%
%
%
%

\vspace*{0.5cm}
\noindent
{\bf The $a$-Euclidean Hopf algebra and the
$\star$-product}

\vspace*{0.3cm}
The operators $\hat{\pat}_\mu$, $M^{rs}$ and
$N^l$ generate transformations on the
coordinate space. In a standard way maps in the coordinate
space can be mapped to maps 
of the space of functions of 
commuting variables.

We first consider the derivatives
\begin{equation}
\label{2.12}
\hat \pat_\mu \to \pat^*_\mu,
\end{equation}
where $\pat^*_\mu$ is the image of the algebraic map $\hat{\partial}_\mu$, 
and as such it is a map of the space of functions of
commuting variables into itself.
In the following, the derivatives
$\partial_\mu$ will always be the ordinary
derivatives $\frac{\partial}{\partial x^\mu}$
on functions of commuting variables.  Such  mappings have previously been discussed
 in \cite{lukrue2}, \cite{local} and \cite{gac2}. 

From the action of  $\hat{\partial}_\mu$ on
symmetric polynomials we can compute the
action of $\partial^*_\mu$ on ordinary functions\footnote{In this form
  first given in \cite{Podles}.}:
\begin{eqnarray}
\label{2.13}
\pat^*_n f(x) & = & \pat_n f(x),\\
\pat_i^* f(x) & = & \pat_i {e^{ia\pat_n} -1\over ia\pat_n} \, f(x).\nonumber
\end{eqnarray}
The derivatives have inherited the Leibniz rule  (\ref{1.8a}):
\begin{eqnarray}
\label{2.14}
\pat_n^*  ( f(x)\star g(x) ) & = &\left(\pat_n^*   f(x)\right) \star g(x)
        + f(x)\star \left( \pat_n^*   g(x) \right),\nonumber\\
\pat^*_i  ( f(x)\star g(x) ) & = & \left( \pat_i^*   f(x) \right) \star g(x) 
+ ( e^{ia \pat_n^*}   f(x) ) \star \left( \pat_i^*   g(x)
        \right).
\end{eqnarray}

We proceed in an analogous way for the generators
$M^{rs}$ and $N^l$. The result is (cp. the momentum representation in
e.g. \cite{lukrue2} and \cite{klm0307})
\begin{eqnarray}
\label{2.15}
N^{*l}   f(x) & = & \Big( x^l\pat_n - x^n\pat_l+ x^l\pat_\mu\pat_\mu { e^{ia \pat_n}-1 \over 2\pat_n } - x^\nu\pat_\nu\pat_l {e^{ia\pat_n} -1- ia\pat_n\over ia\pat_n^2} \Big) f(x),\nonumber\\
\label{2.16}
M^{*rs}   f(x) & = & \left(x^s\pat_r-x^r\pat_s \right) f(x).
\end{eqnarray}
The comultiplication rule (\ref{1.11}) can be reproduced as well
\begin{eqnarray}
\label{2.17}
N^{*l}   \left( f(x)\star g(x)\right) & = & \left( N^{*l}  f(x) \right) \star g(x) +  \left(e^{ia \pat_n^*}   f(x)\right) \star \left( N^{*l}  g(x)\right)\nonumber\\
&& - ia \left( \pat_b^*   f(x) \right) \star \left( M^{*lb}
          g(x)\right),\\
\label{2.18}
M^{*rs} \left(  f(x)\star g(x)\right)&=&( M^{*rs}   f(x)) \star g(x)
        + f(x) \star ( M^{*rs}   g(x)).\nonumber
\end{eqnarray}
The algebra  of functions  with the $\star$-product as multiplication can now be seen 
as a module for the $a$-Euclidean Hopf algebra.

In the previous chapter we have seen that the
Dirac operator $\hat{D}_\mu$ can be interpreted
as a derivative as well. It is natural to
carry it to the algebra of functions with the $\star$-product:
\begin{eqnarray}
\label{2.19}
D^*_n   f(x) & = & \left( {1\over a} \sin (a\pat_n) + 
        {\Delta_{cl}\over ia\pat_n^2}(\cos (a\pat_n) - 1) \right) f(x),\nonumber\\
 \label{2.20}
D^*_i   f(x) & = & \partial_i {e^{-ia\partial_n}-1 \over -ia\partial_n}f(x),
\end{eqnarray}
where $\Delta_{cl}=\pat_i\pat_i$.
The Leibniz rule for the Dirac operator is:
\begin{eqnarray}
D^*_n  ( f(x)\star g(x) ) & = & (D_n^*   f(x)) \star
        (e^{-ia \pat_n^*}   g(x)) + (e^{ia\pat_n^*}   f(x)) \star
        (D_n^*   g(x))\nonumber\\
\label{2.21}
&& + ia
        \left( D_j^* e^{ia\pat_n^*}   f(x) \right) \star
        (D_j^*   g(x)),\\
\label{2.22}
D_i^* (  f(x)\star g(x) ) & = & ( D_i^*   f(x)) \star
        (e^{-ia\pat_n^*}   g(x)) + f(x) \star (D_i^*  
        g(x)).\nonumber
\end{eqnarray}
Finally, the Laplace operator $\hat \Box$:
\begin{equation}
\label{2.23}
\left(\square^*   f(x)\right) = -
{2\over a^2\pat_n^2} (\cos(a\pat_n) - 1) \left( \Delta_{cl}+\pat_n^2
\right) f(x).
\end{equation}

%
%
%
%
%
%

\section{\bf Field equations}

\vspace*{0.5cm}
\noindent
{\bf Fields}

\vspace*{0.3cm}
\noindent
Physical fields are formal power series expansions
in the coordinates and as such elements of the
coordinate algebra:
\begin{equation}
\label{3.1}
\hat \phi(\hat x) = \sum _{ \{\alpha\} } c_{\alpha_1\dots\alpha_n} \, :
(\x^1)^{\alpha_1}\dots (\x^n)^{\alpha_n}:.
\end{equation}
The summation is over a basis in the 
coordinate algebra as indicated by the double points.
The field can also be defined by its coefficient functions
$c_{\{\alpha_1\dots\alpha_n\}}$, once the basis is specified.

Fields can be added, multiplied, differentiated 
and transformed. A transformation is a map
in the algebra and as such can be seen as a map 
of the coefficient functions. We are interested in
the maps that are induced by the  transformations $N_l$:
\begin{equation}
\label{3.2}
\x'^\mu = \x^\mu + \epsilon_l \left( N^l \x^\mu\right).
\end{equation}
The action of $N^l$ on the coordinates  was given
in (\ref{1.4}).
This expresses the transformed coordinate
$\hat{x}'$ in terms of the coordinates $\hat{x}$.
The transformation law of a  scalar field is
usually (for commuting variables) defined as:
\begin{equation}
\label{3.3}
 \phi '(x') = \phi(x).
\end{equation}
Because of the nontrivial coproduct
 of $\hat{N}^l$ operator (\ref{1.11})
acting on the fields (\ref{3.1}) we have to use
\begin{equation}
(1+\epsilon _l\hat{N}^l)\hat{\phi}(\x),
\end{equation}
instead of ${\hat\phi}(\x')$.
Thus, the transformation law of a scalar field  will be defined as follows
\begin{equation}
(1+\epsilon _l\hat{N}^l)\hat{\phi}'(\x)=\hat{\phi}(\x).
\end{equation}

Spinor  fields are defined
analogously:
\begin{equation}
\label{3.5}
(1+\epsilon_l \hat{N}^l)\hat\psi'_\sigma (\x ) = (1+\epsilon_l N^l_{\mbox{\tiny rep}})_{\sigma\rho} \hat
\psi_\rho(\x),
\end{equation}
where $N^l_{\mbox{\tiny rep}}$ is a representation of $N^l$
acting on coordinate independent spinors. The
generalization to vector fields or tensor fields
is obvious.

In the $\star$-product language, we have 
\begin{equation}
\label{3.7a}
\phi'(x)=\phi(x)-\epsilon_l N^{*l} \phi(x),
\end{equation}
where $N^{*l}$ operates  on the coordinates.

The generalization to the operators
$M^{*rs}$ and $\partial^*_\mu$ is straightforward.

\vspace*{0.5cm}
\noindent
{\bf Field equations}

\vspace*{0.3cm}
We introduce
the $a$-deformed Klein-Gordon equation for scalar
fields
\begin{equation}
\label{3.8}
\left( \hat \square + m^2 \right) \hat\phi(\x)=0.
\end{equation}
The invariance of this equation follows from
(\ref{1.12a}) and (\ref{3.3}):
\begin{equation}
\label{3.9}
(1+\epsilon_l \hat{N}^l)\left( \hat \square + m^2 \right) \hat\phi'(\x) = \left( \hat \square + m^2
\right) \hat\phi(\x).
\end{equation}

Similarly the $a$-deformed Dirac equation
\begin{equation}
\label{3.9a}
\left(i \gamma^\la \hat{D}_\la - m\right) \hat\psi(\x) = 0
\end{equation}
 is covariant:
\begin{equation}
\label{3.10}
(1+\epsilon_l \hat{N}^l)\left(i \gamma^\la \hat{D}_\la - m\right) \hat\psi'(\x) = (1+\epsilon_l N^l_{\mbox{\tiny
rep}}) \left( i\gamma^\la \hat{D}_\la-m \right) \hat\psi (\x).
\end{equation}
These equations take the following form in the 
$\star$-formalism:
\begin{equation}
\label{3.11}
\left( \square^* + m^2 \right) \phi(x) = \left( 
        -{2\over a^2 \pat_n^2} (\cos(a\partial_n) - 1) 
        \left( \Delta_{cl}+\partial_n^2 \right) + m^2 \right) \phi(x)
\end{equation}
and
\begin{eqnarray}
\label{3.12}
\left( i\gamma^\la D^*_\la - m \right) \psi(x) & = &
        \Bigg( \gamma^n \Big( {i\over a} \sin (a\pat_n) + {\Delta_{cl}\over 
        a\pat_n^2} (\cos (a\pat_n) - 1) \Big)    \nonumber\\
&&\hspace{2cm}+i\gamma^j\partial_j{e^{-ia\partial_n}-1 \over -ia\partial_n }-m\Bigg) \psi(x).
\end{eqnarray}

We have defined the $a$-deformed Klein-Gordon and Dirac 
equation for fields that are functions of the commuting
variables and  have to be multiplied with the $\star$-product.
These equations are covariant under the $a$-Euclidean transformations.

%
%

\section{The Variational Principle}

We will derive field equations by means of a variational principle
such that the dynamics can be formulated with the help of
the Lagrangian formalism. For this purpose we need an
integral. Algebraically an integral is a linear map of the algebra
into complex numbers:
\begin{equation}
\label{6.1}
\int: \hat{\mathcal{A}}(\hat{x})\longrightarrow \mathbb{C},
\end{equation}

\begin{equation}
\label{6.2}
\int (c_1\hat{f}+c_2\hat{g})= c_1\int\hat{f}+c_2\int\hat{g},\qquad
\forall \hat{f}, \hat{g} \in \hat{\mathcal{A}}(\hat{x}), c_i \in \mathbb{C}.
\end{equation}

In addition we demand the trace property:
\begin{equation}
\label{6.3}
\int\hat{f}\hat{g}=\int\hat{g} \hat{f}. 
\end{equation}
In our case this is essential to define the variational principle.
To find a workable definition of such an integral
we will try to define it in the $\star$-product formalism.
There we can use the usual definition of an integral
of functions of commuting variables. Such an integral
will certainly have the linear property (\ref{6.2}), but in
general it will not have the trace property (\ref{6.3}). It
has, however, been shown in \cite{dietz} that a measure can be introduced
to achieve the trace property: 
\begin{equation}
\label{6.4}
\int\textrm{d}^n x \hspace{1mm}\mu(x)\hspace{1mm}(f(x)\star g(x)) =\int\textrm{d}^n x\hspace{1mm} \mu(x)\hspace{1mm}(g(x)\star f(x)). 
\end{equation}
Note that $\mu(x)$ is not $\star$-multiplied with the other
functions, it is part of the volume element.

It turns out that for the $\star$-product (\ref{2.4a}) and
$\mu(x)$ with the property
\begin{equation}
\label{6.5}
\partial_n \mu(x) =0, \qquad x^j \partial_j \mu (x)= (1-n)\mu(x),
\end{equation}
equation (\ref{6.4}) will be true. This was shown to first order
in $a$ \cite{dietz}, it can be generalized to the full $\star$-
product (\ref{2.4a}) \cite{forthcoming}. 

Technically $\mu(x)$ is needed because $z^j$ occurs in the exponent
of (\ref{2.4a}). Partially integrating, this $z^j$ 
has to be differentiated as well. As $\mu(x)$ has the property
(\ref{6.5}) we find
\begin{equation}
\label{6.6}
\int \textrm{d}^n x \hspace{1mm}\mu(x) f(x) (x^j \p_j g(x)) \rightarrow -\int \textrm{d}^n x \hspace{1mm} \mu(x)  (x^j \p_j f(x)) g(x).
\end{equation}
Expanding the exponent in (\ref{2.4a}) and using
(\ref{6.6}), (\ref{6.4}) can be verified. 

An integral with a measure $\mu(x)$ satisfying (\ref{6.5}) has the
additional property:
\begin{equation}
\label{6.9}
\int\textrm{d}^n x\hspace{1mm} \mu(x)\hspace{1mm}(f(x)\star g(x)) =\int \textrm{d}^n x \hspace{1mm} \mu(x)\hspace{1mm}f(x) g(x). 
\end{equation}

For an arbitrary number of functions multiplied with the 
$\star$-product we can cyclically permute the functions
under the integral
\begin{equation}
\label{6.10}
\int\textrm{d}^n x\hspace{1mm} \mu\hspace{1mm}(f_1\star f_2\star\dots
\star f_k)=\int\textrm{d}^n x\hspace{1mm} \mu\hspace{1mm}(f_k\star
f_1\star f_2\star\dots \star f_{k-1}).
\end{equation}
Thus, any such function can be brought to the 
left or right hand side of the product. For a variation
of some linear combination of such products we always can
bring the function to be varied to one side and then
vary it:
\begin{equation}
\label{6.10a}
\frac{\delta}{\delta g(x)}\int\textrm{d}^n x\hspace{1mm}
\mu\hspace{1mm} f\star g \star h = \frac{\delta}{\delta g(x)}\int\textrm{d}^n x\hspace{1mm}
\mu\hspace{1mm}  g  (h \star f) = \mu\hspace{1mm} h\star f.
\end{equation}

\vspace*{0.5cm}
\noindent
{\bf Hermitean Differential Operators}

\vspace*{0.3cm}

We shall call a differential operator $\mathcal{O}$ hermitean
if 
\begin{equation}
\label{6.15}
\int\textrm{d}^n x\hspace{1mm} \mu\hspace{1mm}\bar{f}\star
\mathcal{O}g =  \int\textrm{d}^n x\hspace{1mm}
\mu\hspace{1mm}\overline{\mathcal{O}f}\star g. 
\end{equation}
It is easy to see that the operators $i\partial^*_i$
or $i D^*_\mu$ are not hermitean by this definition
of hermiticity, though they are by the algebraic
definition (\ref{1.28}). 

Let us first have a look at the differential operator
$\partial^*_i$ as given in  (\ref{2.13}). Due to the property 
(\ref{6.5}) of $\mu$ there is
no problem in partially integrating $\partial_n$.
\begin{eqnarray}
\label{6.17}
\int \textrm{d}^n x\hspace{1mm} \mu\hspace{1mm} \bar{f}\star (\partial_i^* g)& = &
\int \textrm{d}^n x\hspace{1mm} \mu\hspace{1mm} \bar{f}\hspace{1mm}(\partial_i^*
g) =\int \textrm{d}^n x\hspace{1mm} \mu\hspace{1mm} \frac{e^{-ia\p_n}-1}{-ia\p_n}\bar{f}\hspace{1mm} \partial_i g\nonumber\\
& = & -\int \textrm{d}^n x\hspace{1mm} \mu\hspace{1mm} \overline{\partial_i^* f}\hspace{1mm} g-\int \textrm{d}^n x\hspace{1mm} \partial_i\mu\hspace{1mm} \overline{\frac{e^{ia\p_n}-1}{ia\p_n} f}\hspace{1mm}  g
\end{eqnarray}

This is quite similar to the case of polar coordinates in ordinary spacetime
where $i \frac{\partial}{\partial r}$ is not hermitean
due to the measure $r^2\textrm{d}r$, but 
$i(\frac{\partial}{\partial r} +\frac{1}{r})$ is
hermitean. It is tempting to try a similar strategy here. We first define $\rho_i$, which is a logarithmic
derivative of $\mu$ 
\begin{equation}
\label{6.19a}
\rho_i={\partial_i\mu \over 2\mu}. 
\end{equation}
It inherits from $\mu$ the following
properties:
\begin{equation}
\label{6.20}
x^l\partial_l \rho_i=-\rho_i \quad \textrm{and}\quad \partial_n
\rho_i =0.
\end{equation}
Adding $\rho_i$ to $\p_i$ renders a 
derivative $i\tilde{\partial}^*_i$:
\begin{equation}
\label{6.18}
i{\tilde \partial}^*_i = i(\partial_i
+\rho_i)\frac{e^{ia\partial_n}-1}{ia\partial_n}. 
\end{equation}
This $i\tilde{\partial}^*_i$ is hermitean in the sense of (\ref{6.15}):
\begin{equation}
\label{6.21}
\int \textrm{d}^n x\hspace{1mm} \mu\hspace{1mm} \bar{f}\hspace{1mm}\hspace{1mm}i(\partial_i+\rho_i)\frac{e^{ia\p_n}-1}{ia\p_n} g = 
\int \textrm{d}^n x\hspace{1mm} \mu\hspace{1mm} \overline{i(\partial_i+\rho_i)\frac{e^{ia\p_n}-1}{ia\p_n} f}\hspace{1mm}\hspace{1mm}
g.
\end{equation}

The same strategy works for $D^*_\mu$:
\begin{eqnarray}
\label{6.22}
D^*_i&\longrightarrow&{\tilde D}^*_i=(\partial_i
+\rho_i)\frac{e^{-ia\partial_n}-1}{-ia\partial_n},\nonumber\\
D^*_n&\longrightarrow&{\tilde D}^*_n=\frac{1}{ia\partial_n^2}(\partial_k
+\rho_k)(\partial_k
+\rho_k)(\cos(a\partial_n)-1)+\frac{1}{a}\sin(a\partial_n).
\end{eqnarray}
These $i\tilde{D}^*_\mu$
are hermitean in the sense of (\ref{6.15}).

The substitution 
\begin{equation}
\label{6.23}
i\p_i \longrightarrow i(\p_i+\rho_i) =: \pi_i
\end{equation}
does not change the canonical commutation relations,
\begin{equation}
\label{6.24}
[x^j, x^k]=0, \qquad [i\p_i,i\p_l]=0, \qquad  [i\p_i,x^j]=i\de_i^j,
\end{equation}
implies
\begin{equation}
\label{6.25}
[x^j, x^k]=0, \qquad [\pi_i,\pi_l]=0, \qquad  [\pi_i,x^j]=i\de_i^j,
\end{equation}
and vice versa. This can be seen by making use of
the properties (\ref{6.20}) of $\rho_i$.

Thus, replacing $i\partial_i$ by $\pi_i$ does not
change the algebraic properties of the differential
operators. This suggests to introduce $\tilde{M}^{*rs}$
and $\tilde{N}^{*l}$ as well. These operators will satisfy
the same commutation relations  as 
$M^{*rs}$, $N^{*l}$, $\p_\mu^*$ and $D^*_\mu$. 
In the sense of (\ref{6.15}) the operators 
$i\tilde{M}^{*rs}$ will be hermitean, $i\tilde{N}^{*l}$ not.

A proper action for a spinor field $\tilde{\psi}$ would be:
\begin{equation}
\label{6.26}
S=\int\textrm{d}^n x\hspace{1mm} \mu\hspace{1mm} \overline{\tilde{\psi}}\star(i\gamma^\lambda \tilde{D}^*_\lambda -m)\tilde{\psi}.
\end{equation}
By varying with respect to $\overline{\tilde{\psi}}$ we
obtain
\begin{equation}
\label{6.27}
\mu (i\gamma^\lambda \tilde{D}^*_\lambda -m)\tilde{\psi} = 0.
\end{equation}
Guided by the example of polar coordinates we compute
\begin{equation}
\label{6.28}
\tilde{D}^*_i \mu^\al = \mu^\al(\p_i+(2\al+1)\rho_i)\frac{e^{ia\p_n}-1}{ia\p_n},
\end{equation}
and similar for $\tilde{D}^*_n$.
If we choose $\al= - 1/2$ we obtain 
\begin{equation}
\label{6.29}
\tilde{D}^*_\lambda \mu^{-\frac{1}{2}} = \mu^{-\frac{1}{2}}D^*_\lambda.
\end{equation}
This suggests the introduction of the field
\begin{equation}
\label{6.30}
\tilde{\psi}=\mu^{-\frac{1}{2}}\psi.
\end{equation}
The field $\psi$ satisfies the Dirac equation
as it was introduced in (\ref{3.12}).
\begin{equation}
\label{6.31}
(i\gamma^\lambda D^*_\lambda -m)\psi=0.
\end{equation}
This equation can be derived from the action
\begin{equation}
\label{6.14}
S=\int\textrm{d}^n x\hspace{1mm} \overline{\psi}(i\gamma^\lambda D^*_\lambda -m)\psi,
\end{equation}
which is exactly the action we obtain by substituting
$\tilde{\psi}\rightarrow\mu^{-\frac{1}{2}}\psi$ in the action (\ref{6.26}), after dropping the
$\star$ from the integral with the help of $\mu$.

\section*{Acknowledgment}
We thank Frank Meyer for his help and many useful discussions.

L. J.  gratefully acknowledges the support of the Alexander
von Humboldt Foundation and the Ministry of Science and Technology
of the Republic of Croatia under the contract 0098003.

M. D.  gratefully acknowledges the support of the Deutscher
Akademischer Austauschdienst.

\end{document}